\documentstyle[prl,floats,aps]{revtex}

\begin{document}

\draft

\twocolumn[\hsize\textwidth\columnwidth\hsize\csname @twocolumnfalse\endcsname

\title{Pathologies of hyperbolic gauges in general relativity \\
and other field theories}
\author{Miguel Alcubierre and Joan Mass\'{o}}

\address{Max-Planck-Institut f\"{u}r Gravitationsphysik
(Albert-Einstein-Institut) \\ Schlaatzweg~1, D-14473 Potsdam, Germany.}
\date{December 9, 1997}
\maketitle

\begin{abstract}{We present a mathematical characterization of hyperbolic gauge
    pathologies in electrodynamics and general relativity. We show
    analytically how non-linear gauge terms can produce a blow-up of
    some fields along characteristics. We expect similar phenomena to
    appear in any other gauge field theory. We also present numerical
    simulations where such blow-ups develop and show how they can be
    properly identified by performing a convergence analysis. We
    stress the importance of these results for the particular case of
    numerical relativity, where we offer some cures based on the use
    of non-hyperbolic gauges.}
\end{abstract}


\vskip2pc]

\narrowtext

\input{epsf.sty}

%%%%%%%%%%%%%%%%%%%%
%%%  MAIN  TEXT  %%%
%%%%%%%%%%%%%%%%%%%%

During the last decades, gauge field theories have become the paradigm
of fundamental physics.  In such theories, gauge invariance implies
that the `physics' is independent of our choice of gauge.  Yet in some
cases, of which general relativity (GR) is a very good example, it can
be difficult to separate the gauge from the physical degrees of
freedom.  This is of particular importance in numerical simulations of
non-linear field theories where one needs to distinguish physical
singularities from pure gauge pathologies.  In such cases, exact
solutions are not available, and the simple idea of a change of gauge
can become a daunting task.  Because of this, understanding the
properties of particular gauge choices becomes an issue of great
relevance.

In a recent paper\cite{Alcubierre97} it was shown, to the surprise of
many in the numerical relativity community, that certain gauge choices
(throughout we use the term `gauge' applied to relativity to mean only
slicing conditions) in relatively simple scenarios could lead to the
development of what were there called `coordinate shocks'.  Here we
will show that the term `shock' used in that paper is misleading and
will refer to them instead as gauge pathologies.  The recent
Bona-Mass\'o hyperbolic formulation of the Einstein Equations
\cite{BonaMasso} allowed the author of reference\cite{Alcubierre97} to
study the structure of the gauge condition and how its non-linearity
could cause gauge pathologies. It was shown numerically that these
pathologies seemed to occur, but no proof that they were real
discontinuities was given.  This work left some basic questions
unanswered: How can we characterize these pathologies mathematically
and numerically? How generic are they? Do similar phenomena occur in
other gauge theories?

Here we give for the first time a mathematical characterization of
hyperbolic gauge pathologies (where by hyperbolic gauge we mean choices
where the gauge is evolved using a hyperbolic equation) based on the
theory of nonlinear waves\cite{Kichenassamy}.  We focus our attention
on two cases: electrodynamics (ED) and spherically symmetric GR.  We
analyze the structure of the equations and show how non-linear gauge
terms can produce a blow-up of some fields along
characteristics\cite{Kichenassamy}.  Such a blow-up indicates a gauge
singularity and corresponds to the `coordinate shocks' of
reference\cite{Alcubierre97}.  The blow-ups described here are much
stronger singularities than shocks since some of the propagating
fields become in fact infinite.  We also show how the blow-ups can be
identified numerically by analyzing the convergence properties of
computational simulations.  Finally, we propose some ``cures'' for
these pathologies based on changing the hyperbolic nature of the gauge
condition.  Thought we only study the cases of ED and GR, we expect
similar phenomena to appear in any other gauge field theory.

{\em Electrodynamics}--- Surprisingly, gauge effects can produce
blow-ups along characteristics even in simple systems such as ED. We
are not aware of any previous analysis of gauge pathologies in ED,
probably because well behaved gauge choices are intuitively clear
there. In the following, we will follow closely the notation of
Refs.\cite{Alcubierre97,BonaMasso}.

We will write the equations of ED as a first order initial value
problem in the following way:
\begin{mathletters} \label{eq:ED2} \begin{eqnarray}
\partial_t A_i &=& - \left( E_i + \psi_i \right) \;\; , \\
\partial_t D_{ij} &=& - \partial_i \left( E_j + \psi_j \right)
\;\; , \\
\partial_t E_i &=& \partial_i {\rm tr} D - \textstyle{\sum_j}
\partial_j D_{ji} \;\; ,
\end{eqnarray} \end{mathletters}
where $\phi$ and $\vec{A}$ are the scalar and vector potentials,
$\vec{E}$ the electric field, and where we have introduced the
quantities $D_{ij} := \partial_i A_j$ and $\psi_i := \partial_i \phi$.
We also have one constraint which in vacuum takes the form $\nabla
\cdot \vec{E} = 0$.

We are now free to decide how $\{\phi,\psi_i\}$ will evolve, {\em i.e.}
we are free to choose the gauge.  Here we make a choice similar to the one
we will make later in relativity
\begin{equation}
\partial_t \phi = - f(\phi) \, {\rm tr} D  \; ,
\label{eq:gaugeED}
\end{equation}
with $f(\phi)>0$ but otherwise arbitrary ($f=1$ corresponds to the
familiar Lorentz gauge).  It is through the gauge function $f$ that we
introduce a non-linearity into ED, which is otherwise a completely
linear theory.

The resulting system of equations can be shown to be hyperbolic (in
the sense of having a complete family of eigenfields).  For a given
direction $x$ we find the characteristic structure: 4 fields propagate
along the physical light cones (speed $\pm 1$)
\begin{equation}
{\omega_p}_\pm := \left( E_p + \psi_p \right) \pm D_{xp} \;\; ,
\qquad p \neq x \;\; ,
\end{equation}
and 2 fields propagate with the `gauge speeds' \,$\pm\, \sqrt{f}$
\begin{equation}
{\omega_g}_\pm := \psi_x \pm \sqrt{f} \, {\rm tr} D \;\; . 
\end{equation}
The remaining 13 fields move along the time lines.

Let us now define ${\Omega_g}_\pm := \sqrt{f} \, {\omega_g}_\pm$.
We find
\begin{equation}
\partial_t {\Omega_g}_{\pm} \pm
\sqrt{f} \, \partial_x {\Omega_g}_{\pm} = \pm \, \frac{f'}{4\,f}
\left({\Omega_g}_{\pm}^2 + {\Omega_g}_{+}{\Omega_g}_{-} \right)
\end{equation}
For $f'\neq0$, the quadratic term in ${\Omega_g}_\pm$ can produce a
blow-up along a characteristic in a finite time.  We can predict the
time when this will happen if we restrict ourselves to the case
${\Omega_g}_{-}=0$, and take $f=e^{a \phi}$.  Along a characteristic
we will have
\begin{equation}
\frac{d {\Omega_g}_{+}}{dt} = \frac{a}{4} \, {\Omega_g}_{+}^2
\;\; ,
\end{equation}
which can be easily integrated to find
\begin{equation}
{\Omega_g}_+ = \frac{\Omega_0}{1 - a \Omega_0 t / 4} \;\; .
\end{equation}

For \mbox{$a\,\Omega_0 > 0$}, a gauge pathology (a blow-up of
${\Omega_g}_+$) will appear at a finite time $T$ given by
\begin{equation}
T = 4 / a \Omega_0 \;\; .
\end{equation}

\begin{figure} \def\epsfsize#1#2{0.55#1}
\centerline{\epsfbox{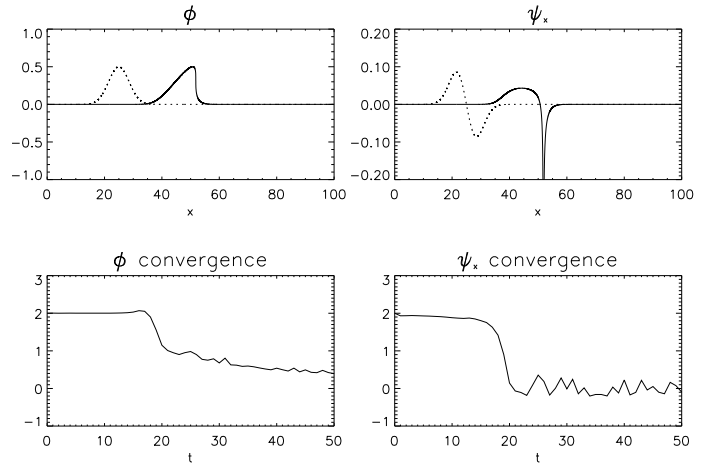}}
\caption{Numerical simulation of a gauge wave in electrodynamics.
  The first two plots show the initial (dotted line) and final (solid
  line) values at $t=20$ of the scalar potential $\phi$ and its
  derivative $\psi_x$.  The lower plots show the convergence rates as
  a function of time.}
\label{fig:ED} \end{figure}

We have constructed a numerical code to evolve this system of
equations and reproduce this blow-up.  Fig.~\ref{fig:ED} shows the
results of a one-dimensional simulation for which a shock is expected
at $T=19.92$.  The first two plots show the initial (dotted line) and
final (solid line) values at $t=20$ of $\phi$ and $\psi_x$.  A sharp
gradient has developed in $\phi$, while a large spike has appeared in
$\psi_x$.

Now, if we had not performed the mathematical analysis of the system,
we could ask ourselves how can we know that these features correspond
to a real blow-up and not to the development of large but smooth
gradients.  After all, nonlinear systems may develop features that are
difficult to resolve numerically.  As an example, one of the key
successes of numerical relativity has been the discovery of critical
phenomena\cite{ChoptuikPRL}.  Confidence in these results required
advanced adaptive mesh refinement and the careful use of convergence
tests\cite{ChoptuikCONV}.  Here we have performed similar convergence
studies: we consider a series of different resolutions and find the
rate at which a global measure of their relative errors converges to
zero (the `self-consistent' convergence rate).  Our code uses a second
order accurate scheme, so we expect the convergence rate of our
solutions to be 2.  The lower two plots of Fig.\ref{fig:ED} show the
convergence rates for $\phi$ and $\psi_x$ obtained from runs at three
resolutions (5000, 10000 and 20000 points).  As expected, for $t<20$
the convergence rate is close to 2, while after that it drops
dramatically indicating loss of convergence.  Moreover, the spikes in
$\psi_x$ become larger and larger with increased resolution.  We then
conclude that they correspond to a real infinity and not just a large
gradient.  The lesson is clear: we can characterize numerically the
appearance of a blow-up using convergence analysis.

It is important to consider the effect that the choice of finite
difference method has on the results described above.  When we used
standard methods like a leap frog scheme, the simulations crashed very
soon after the spike in $\psi_x$ started to develop.  Using instead
shock capturing\cite{Leveque} techniques we were able to follow the
evolution much further, which allowed us to determine more precisely
the time of the blow-up.  Notice that this further evolution is
non-physical: the shock capturing techniques help to maintain the
simulation stable but this is of no physical relevance since we the
true solution is not really a shock wave.  A real infinity has
developed and no numerical method will converge after that.

{\em Spherically symmetric general relativity}-- Let us now consider
the case of GR in spherical symmetry.  The basic variables are the
lapse function $\alpha$, the spatial metric components
$\{g_{rr},g_{\theta\theta}\}$ and the extrinsic curvature components
$\{K_{rr},K_{\theta\theta}\}$.  Since we are interested in hyperbolic
gauges we will use the following gauge condition~\cite{BonaMasso}
\begin{equation}
\partial_t \alpha = - \alpha^2\,f(\alpha) \, {\rm tr} K \;\; ,
\label{eq:gaugeBM} \end{equation}
with $f(\alpha) > 0$ but otherwise arbitrary (\mbox{$f = 1$} now
corresponds to harmonic slicing).  It can be shown that with this
gauge condition the evolution equations of GR can be written in first
order hyperbolic form\cite{BonaMasso}.

The particular form of the equations for spherical symmetry and its
characteristic structure can be found in\cite{Alcubierre97}.  There it
was shown that there are two families of travelling modes, one that
moves with the speed of light, and one that moves with a `gauge speed'
that depends on the value of the function $f$ and that reduces to the
speed of light for harmonic slicing.  In general, both types of
travelling modes can develop gauge pathologies.  Here we will
concentrate in the particular case of harmonic slicing ($f = 1$) where
only one type of pathology appears.  Note that this gauge is the
preferred choice for many of the new hyperbolic formulations of the
Einstein equations\cite{OtherHyp}.  The analysis of the system for
other forms of $f$ will be considered elsewhere\cite{preparation}.

We will start by defining
\begin{equation}
\Omega_\pm := \alpha / g_{\theta\theta} \left( K_{\theta\theta}
\,\pm\, D_{r\theta\theta} / g_{rr}^{1/2} \right) \;\; ,
\end{equation}
where \mbox{$D_{r\theta\theta} := 1/2 \; \partial_r g_{\theta\theta}$}.
>From the form of evolution equations given in\cite{Alcubierre97} it is
easy to show that, for \mbox{$f=1$}, we will have
\begin{equation}
\partial_t \Omega_\pm \pm \alpha/g_{rr}^{1/2} \, \partial_r \Omega_\pm
= \Omega_+ \Omega_- - \Omega_\pm^2 + \alpha^2/g_{\theta\theta} \;\; .
\label{evOmega}
\end{equation}
We see that the $\Omega_\pm$ represent outgoing \mbox{(+)} and ingoing
\mbox{(-)} modes travelling with the speed of light.  Unfortunately,
the last term on the right hand side of the above equations makes it
impossible to separate them, and so prevents one from predicting when
a blow-up caused by the quadratic source terms might occur.
Nevertheless, one can see that the quadratic term in $\Omega_\pm$
appearing in~(\ref{evOmega}) is only dangerous for negative values of
$\Omega_\pm$.  Moreover, it is easy to see that if $\Omega_\pm$ is
initially positive it will not change sign.  Now, if our initial data
is time-symmetric ($K_{ij}=0$) and such that the metric function
$g_{\theta\theta}$ is monotonic, then we will have $\pm \, \Omega_\pm
> 0$.  This means that no blow-ups can develop for outgoing modes:
Only ingoing modes can produce a gauge pathology.  Of course, whether
they will or not depends on the precise form of the initial data.

In reference\cite{Alcubierre97} it was shown that pathologies appeared for a
particular choice of initial data in a black hole spacetime.  Here we
give another example of how pathologies can form from apparently
simple initial data.  We choose the standard initial data for the metric
and extrinsic curvature that has been successfully used for most black
hole simulations to date \cite{Bernstein,BonaMassoStela,3Dncsa}.  It
corresponds to an isotropically sliced Schwarzschild black hole with
time symmetry.  The key difference is that, instead of choosing an
initial lapse that satisfies the maximal slicing condition, we choose
the following `Gaussian' profile:
\begin{equation}
\alpha = 1 - A \, \exp \left[ ((r-r_0)/\sigma)^p \right] \;\; .
\end{equation}

We have used 3 independent numerical codes to evolve the system, one
based on the standard Arnowitt-Deser-Misner (ADM)
formulation\cite{ADM} that uses a simple leap frog scheme, and two
based on the Bona-Mass\'{o} hyperbolic formulation\cite{BonaMasso}
that use shock capturing methods\cite{Leveque}.  All three codes
produce similar results with one important difference: the ADM code
crashes soon after the pathologies start to develop, while the other
two codes are capable of continuing past this point.

\begin{figure}[ht] \def\epsfsize#1#2{0.55#1}
\centerline{\epsfbox{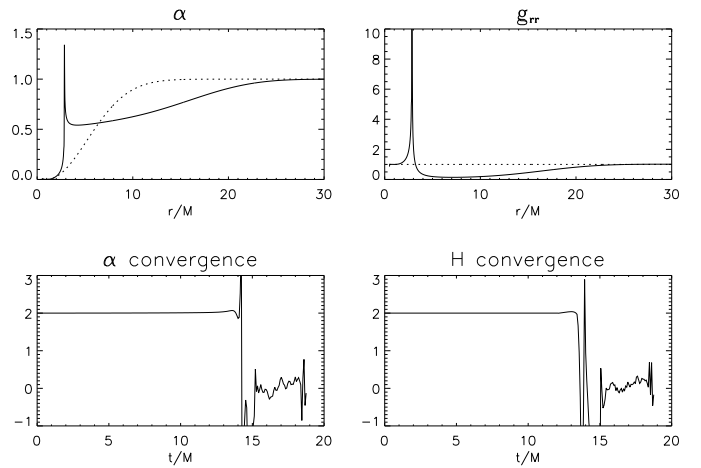}}
\caption{Numerical simulation of a spherically symmetric
black hole spacetime. The first two plots show the initial (dotted
line) and final (solid line) values at $t=15M$ of the lapse $\alpha$
and the conformal metric function $g_{rr}$. The lower plots show the
convergence rates for the lapse and the hamiltonian constraint.}
\label{fig:BH} \end{figure}

We report the results of a particular simulation obtained using one of
the hyperbolic codes.  The first plot in Fig.\ref{fig:BH} shows the
initial (dotted line) and final (solid line) values of the lapse at
time $t=15M$ for the case $\{A=1, r_0=M/2, \sigma=6M, p=2\}$, with $M$
the mass of the black hole.  The next plot shows the same for the
conformal metric function $g_{rr}$.  Notice how both the lapse and the
radial metric develop large spikes, the sizes of which increase with
resolution.  The bottom plots show the global convergence rate of the
lapse and the hamiltonian constraint obtained from runs at 4000, 8000
and 16000 grid points, with the outer boundary located at $40M$.
Clearly, we have a gauge pathology, whose blow-up time seems to be
$t=(14 \pm 1)M$.

Although the simulation presented here corresponds to harmonic
slicing, gauge pathologies also develop with other hyperbolic gauge
choices. An important difference is that in the harmonic case the
lapse becomes infinite, which indicates that the time slicing becomes
null.  In the other cases, the lapse becomes discontinuous but remains
finite, and the time slicing develops a kink instead.  We have
performed many simulations studying the parameter space
$\{A,\sigma,p\}$ and found similar results.  Details of all these
studies will be given elsewhere\cite{preparation}.

Another crucial aspect of the problem is that of the numerical
resolution of the gauge pathologies.  We have found that if we evolve
the previous system with only 200 grid points, the pathologies do not
seem to form.  The lapse grows until a certain value is reached and
then it propagates out in a smeared manner due to the large numerical
viscosity. The solution ``looks good'', although it is non-physical,
as a proper convergence test reveals.

The impact of these results on three dimensional (3D) numerical
relativity should not be underestimated.  Note that in the previous
examples we have used thousands of points to be able to show very
sharp fall-offs of the convergence rate and have a good estimate of
the blow-up time.  Even if this sharpness will not be possible at the
resolutions currently available for 3D computations, at medium and low
resolutions one can already see that the convergence fails at late
times.

We should stress again the fact that the development of these pathologies
depends crucially on the form of the initial data.  For different
choices of the initial lapse function, one can find that harmonic
slicing is perfectly well behaved, as the simulations presented in
reference~\cite{BonaMassoStela} show.  In fact, one can even find
explicitely a harmonic slicing of a black hole in which the metric is
static~\cite{BonaMasso88}.

{\em ``Cures''}-- The appearance of gauge pathologies might seem to
put into question the practical value of non-linear hyperbolic gauges
in numerical studies of gauge field theories.  After all, in a general
situation it might be very difficult to know {\em a priori}\/ if our
initial data will develop such a pathology.

For ED, the solution is clear: use a gauge that decouples the
characteristic speeds from the dynamics, {\em i.e.} use the Lorentz
gauge.  Unfortunately, this will not work in relativity where the
characteristic speeds cannot be decoupled from the dynamics.  We can
think of at least two different ways to solve the problem.  Both
involve changing the character of the equations for the {\em gauge
  only}: all other equations (the `physics') remain hyperbolic.

The first approach implies using an elliptic gauge condition of which
maximal slicing is the best known example.  One can then either use an
elliptic gauge always or, in cases where it might be of interest, use
a hyperbolic gauge for some time and then switch to an elliptic gauge
when a pathology is about to form.  We have tested this idea and found
that it works very well in practice\cite{preparation}.

The second approach consists of adding dissipation to our gauge
condition.  We will then have a parabolic equation and intuition tells
us that this should prevent the pathologies.  We then propose the
gauge condition
\begin{equation}
\partial_t \alpha = - \alpha^2 \left[
f(\alpha) \, {\rm tr} K - \xi(\alpha) \,
\nabla^2 \alpha \right] \;\; ,
\end{equation}
with $f,\xi>0$ but otherwise arbitrary.  Notice that the numerical
treatment of the diffusion term requires either a very stringent
Courant condition or the use of implicit techniques.  Note also that
this term should be kept with the same coefficient for all resolutions
of a given simulation, as it does not correspond to a simple
``artificial viscosity''\cite{3Dncsa}, but rather to an explicit
change of the character of the gauge condition.  We have tried this
condition in spherically symmetric GR and found that it also prevents
the development of gauge pathologies\cite{preparation}.

In conclusion, we have presented for the first time a characterization
of hyperbolic gauge pathologies in ED and GR.  We have shown how the
coupling of characteristic gauge speeds to the dynamics produces a
nonlinear blow-up mechanism, and how a careful convergence analysis
can indicate the appearance of such a blow-up in numerical
simulations.  The origin of these pathologies is in the finite speed
of propagation of the gauge modes, and therefore a way to avoid them
is the use of elliptic or parabolic gauges with infinite gauge speeds.

Thought we have concentrated in the cases of ED and GR, we expect
similar pathologies to arise in any other gauge field theory.  Because
of this we feel that further mathematical study of these phenomena
will be of fundamental importance for future numerical simulations of
non-linear field theories.

We would like to thank Gabrielle Allen, Carles Bona, Bernd
Br\"{u}gmann, Carsten Gundlach, Ed Seidel, Joan Stela, Wai-Mo Suen and
Paul Walker for helpful discussions.  Special thanks to Bernard Schutz
and Richard Matzner for a very careful and critical revision of the
original manuscript.

%%%%%%%%%%%%%%%%%%%%
%%%  REFERENCES  %%%
%%%%%%%%%%%%%%%%%%%%

%%%%%%%%%%%%%%%%%
%%%  FIGURES  %%%
%%%%%%%%%%%%%%%%%

%%%%%%%%%%%%%%%%%%%%%%%
%%%  END  DOCUMENT  %%%
%%%%%%%%%%%%%%%%%%%%%%%

\end{document}